\documentclass[iop]{emulateapj}
\usepackage{epstopdf}

\newcommand{\psiztwo}{1.1}

\slugcomment{Accepted for ApJ}

\shorttitle{Spin Evolution: Effect of Accretion-Powered Stellar Winds}
\shortauthors{Matt et al.}

\begin{document}

\title{Spin Evolution of Accreting Young Stars. II. Effect of Accretion-Powered Stellar Winds}

\author{Sean P. Matt$^{1,2}$,
        Giovanni Pinz\'on$^{3}$,
        Thomas P. Greene$^{2}$, and
        Ralph E. Pudritz$^{4}$}

\affil{$^1$Laboratoire AIM Paris-Saclay, CEA/Irfu Universit\'e
  Paris-Diderot CNRS/INSU, 91191 Gif-sur-Yvette, 
France; sean.matt@cea.fr}

\affil{$^2$NASA Ames Research Center, M.S. 245-6, Moffett Field, CA
94035-1000, USA; thomas.p.greene@nasa.gov}


\affil{$^3$Observatorio Astron\'omico Nacional, Facultad de Ciencias,
  Universidad Nacional de Colombia, Bogot\'a, Colombia;
  gapinzone@unal.edu.co}


\affil{$^4$Physics \& Astronomy Department, McMaster University,
  Hamilton ON, Canada L8S 4M1, pudritz@physics.mcmaster.ca}

\begin{abstract}

  We present a model for the rotational evolution of a young,
  solar-mass star interacting magnetically with an accretion disk.  As
  in a previous paper (Paper~I), the model includes changes in the
  star's mass and radius as it descends the Hayashi track, a
  decreasing accretion rate, and a prescription for the angular
  momentum transfer between the star and disk.  Paper~I concluded
  that, for the relatively strong magnetic coupling expected in real
  systems, additional processes are necessary to explain the existence
  of slowly rotating pre-main-sequence stars.  In the present paper,
  we extend the stellar spin model to include the effect of a
  spin-down torque that arises from an accretion-powered stellar wind.
  For a range of magnetic field strengths, accretion rates, initial
  spin rates, and mass outflow rates, the modeled stars exhibit
  rotation periods within the range of 1--10 days in the age range of
  1--3 Myr.  This range coincides with the bulk of the observed
  rotation periods, with the slow rotators corresponding to stars with
  the lowest accretion rates, strongest magnetic fields, and/or
  highest stellar wind mass outflow rates.  We also make a direct,
  quantitative comparison between the accretion-powered stellar wind
  scenario and the two types of disk-locking models (namely the X-wind
  and Ghosh \& Lamb type models) and identify some remaining
  theoretical issues for understanding young star spins.

\end{abstract}


\section{Introduction} \label{sec_intro}

There remain a number of unexplained phenomena related to the spin
rates of young, low mass ($\la 2 M_\odot$) stars
\citep[e.g.,][]{herbstea07, scholz09, irwinbouvier09}.  At the
youngest stages, the fundamental dilemma is that a large fraction of
stars rotate more slowly than expected, given that the stars are in
the process of contracting and are often accreting material with high
specific angular momentum (i.e., from a circumstellar disk).  In the
absence of significant spin-down torques, one would expect these stars
to spin near breakup speed.  However, observations reveal that
approximately half of solar mass stars with ages of less than a few
million years rotate at less than $\sim$10\% of their breakup
velocities \citep[e.g.,][]{rebull3ea04, herbstea07, scholz09}.  Thus,
for a large fraction of pre-main-sequence stars, there exists some
mechanism that removes significant amounts of angular momentum.

In a previous paper \citep[][hereafter Paper~I]{mattea10}, we examined
whether the magnetic connection between a young star and surrounding
accretion disk can be solely responsible for the angular momentum
loss.  That work employed a model similar to some previous studies in
the literature \citep[][]{cameroncampbell93, yi94, yi95, cameron3ea95,
  armitageclarke96}, which computes the time-evolution of the mass,
radius, and rotation rate of an accreting star.  Specifically, Paper~I
presented calculations of the spin rate of a one-solar-mass star as it
evolves from an age of 30,000 yr to 3Myr.  The calculations considered
the contraction of the star during the \citet{Hayashi:1961p3236} phase
and an exponentially decreasing accretion rate.  The calculations also
considered a range of initial accretion rates, spin rates, and
magnetic field strengths, approximately representative of typical
observed or expected values for low mass pre-main-sequence stars.
Rather than attempt to explain all phenomena related to young star
spins, the primary goal in Paper~I (and in the present paper) was to
address the fundamental question of whether the models could produce
spin rates within the observed typical range of $\sim1$--10 days.

The main difference with previous works is that the model in Paper~I
used the magnetic torque formulation presented by
\citet{mattpudritz05}, which includes the effects of the magnetic
coupling (diffusivity) in the disk and the loss of magnetic connection
that occurs when field lines are sufficiently twisted azimuthally
\citep[e.g.,][]{uzdensky3ea02}.  When considering the strong magnetic
coupling (i.e., low diffusivity or high magnetic Reynolds number)
expected in real systems, Paper~I concluded that the spin-down torques
arising from the star-disk interaction alone were not sufficient to
produce rotation periods longer than $\sim 3$ days at ages of 1--3
Myr.  The main result of Paper~I suggests that additional spin-down
torques are necessary to explain the slow rotators.  That model
neglected any torques that may arise along open field regions, such as
torques from stellar winds, so a natural next step is to consider
whether or under which circumstances such torques may be important.

Powerful and large-scale jets and outflows are observed to emanate
from accreting young stars with outflow rates of the order of
0.01--0.1 times the accretion rates \citep[e.g.,][]{reipurthbally01,
  cabrit07}.  Stellar winds---that is, outflows that are magnetically
connected to the star, as opposed to the accretion disk---may be an
important component in these large-scale flows
\citep[e.g.,][]{Decampli:1981p2972, Kwan:1988p2882, Fendt:1995p2893,
  Fendt:1996p2895, Paatz:1996p2904, hiroseea97, fendtelstner00,
  Bogovalov:2001p2926, Sauty:2002p2932, Sauty:2004p2940,
  Meliani:2006p2883, Matsakos:2008p2936, Matsakos:2009p2937, fendt09,
  romanovaea09, Sauty:2011p2934}.  Furthermore, high-resolution
spectroscopic observations of these systems show some emmision line
features that appear to be best explained by the presence of stellar
winds, and that these signatures generally correlate with the
accretion rates \citep[e.g.,][]{Hartmann:1982p2885, Kwan:1988p2882,
  Hartmann:1990p2900, edwardsea03, Dupree:2005p3006, edwardsea06,
  kurosawa3ea06, kwan3ea07, JohnsKrull:2007p3020, kurosawa3ea11}. The
idea that powerful stellar winds could also be the primary agent for
removing angular momentum from the star has been suggested or explored
by several authors \citep{Hartmann:1982p2888, Mestel:1984p2919,
  Hartmann:1989p2909, Tout:1992p1467, Paatz:1996p2904, hiroseea97,
  Ferreira:2000p1706, VonRekowski:2004p2874, romanovaea05,
  mattpudritz05l, VonRekowski:2006p2879, mattpudritz08III,
  romanovaea09, zanniferreira09, Sauty:2011p2934}.
\citet{mattpudritz05l, mattpudritz08III} showed that a stellar wind
will be much more effective at spinning down a star than the magnetic
connection between the star and disk, as long as the mass outflow rate
is high enough.  In order to maintain a sufficiently high mass loss
rate, \citet{mattpudritz05l} suggested that the stellar wind is
somehow powered by the accretion process, and \citet{mattpudritz08II}
suggested that accretion onto the star may power a stellar wind by the
excitation of a large flux of Alfv\'en waves along the open field
lines.  The idea of a wave-driven and accretion-powered stellar wind
was further explored by \citet{Cranmer:2008p1657,
  Cranmer:2009p1647}.

In the present paper, we explore the role of stellar winds in the spin
evolution of a young star that is also interacting magnetically with
an accretion disk.  The model (described in \S \ref{sec_model}) is
identical to that of Paper~I except that the star experiences an
additional torque from a stellar wind.  This introduces only one new
parameter, the mass outflow rate in the wind, which we take to be a
fraction of the accretion rate, as expected for an accretion-powered
stellar wind (hereafter, APSW).  The results in section
\ref{sec_results} demonstrate how an APSW, acting during the entire
accretion history, may explain the typical range of observed spin
rates.  In section \ref{sec_comparison}, we compare the present model
with disk-locking models.  A discussion of the conclusions of this
work and remaining challenges are contained in section
\ref{sec_conclusions}.



\section{Stellar Spin Evolution Model} \label{sec_model}

In order to best compare the effects of powerful stellar winds to the
torques arising in the star-disk magnetic interaction alone, we employ
the same assumptions, equations, and method as Paper~I, except that we
have included the mass loss and torque resulting from the wind.  This
section lists the relevant equations and points out where there are
differences from Paper~I (where the reader will also find more
details).

     \subsection{Mass Flow} \label{sec_mdot}

The mass accretion rate onto the stellar surface follows an
exponential decay,
\begin{equation}
\label{eq_mdot}
%
\dot{M_{a}} = \dot M_{a0} \;\; e^{-t/t_a},
\end{equation}
where $\dot M_{a0}$ is the mass loss rate at $t=0$, and $t_a$ is the
exponential decay timescale.  Following Paper~I, we adopt $t_a = 10^6$
yr, for all models, and two different values of $\dot M_{a0}$ so that
the models span a range of accretion rates.  The two values correspond
to a ``low'' accretion rate with $\dot M_{a0}=10^{-8} M_\odot$ yr$^{-1}$ and
a ``high'' accretion rate with $\dot M_{a0}=10^{-7} M_\odot$ yr$^{-1}$ (see
Fig.\ 1 of Paper~I\footnote{Paper~I employed a different nomenclature,
  where $\dot M_{a0} \equiv M_D t_a^{-1}$, but regardless we adopt the same
  two values for the initial accretion rates as in Paper~I.}). In all
cases, we chose an initial stellar mass, such that the mass of the
star at the end of the computation (at an age of 3~Myr) equals one
solar mass (see Fig.\ 2 of Paper~I).

At the same time, the present work considers the additional effect of
a stellar wind.  The general picture of simultaneous inflow and
outflow from the star is as follows.  A fraction of the stellar
magnetic flux connects to the accretion disk, which is responsible for
the magnetic torques between the two and for channeling an accretion
flow onto the star.  The stellar wind flows from the polar regions of
the star, where the magnetic field lines reach beyond the
magnetosphere.  This region of the magnetic field is susceptible to
being opened and kept open by energetic processes, such as disk and/or
stellar winds \citep[e.g.,][]{safier98, mattpudritz05} or differential
rotation between the star and disk \citep[e.g.,][]{uzdensky3ea02}.
The basic idea behind APSWs is that a fraction of the accretion energy
that is dissipated near the surface of the star ultimately powers a
stellar wind.  For this to work, energy deposited along the accreting
field lines should transfer laterally (likely via waves) across the
accretion flow and into the open field region.  In the open field
region, this energy then drives an enhanced stellar wind, where the
outflowing mass is from the star and is distinct from the material
that is currently accreting onto the star.  A more detailed
description of these processes are given in (e.g.)\
\citet{mattpudritz05l} and \citet{Cranmer:2008p1657}, and an
illustration of the model is given in Figure 1 of each of those works.

When the energy responsible for driving the wind ultimately comes from
the accretion process, it stands to reason that the mass outflow rate
$\dot M_{\rm w}$ will be ultimately tied to the accretion rate
\citep[][]{mattpudritz05, Cranmer:2008p1657, Cranmer:2009p1647}.
Thus, we parameterize the stellar wind outflow rate using
\begin{eqnarray}
\label{eq_swfrac}
\chi \equiv {\dot M_w \over \dot M_a}.
\end{eqnarray}
The evolution of the star's mass is thus given by
\begin{eqnarray}
\label{eq_mdotstar}
\frac{dM_* }{dt} = (1-\chi) \dot M_a.
\end{eqnarray}

\citet[][]{Hartmann:1989p2909} suggested that a value of $\chi\sim0.1$
is necessary for the angular momentum outflow rate in a stellar wind
to be comparable to the angular momentum deposited onto the star from
accretion of disk material \citep[and see][]{mattpudritz05}.
Observations of large-scale outflows from T Tauri stars and
(presumably younger) Class I sources indicate mass outflow rates of
$\sim 0.1$ times the accretion rates, with about an order of magnitude
of scatter and/or uncertainty in this value \citep[e.g., see review
by][]{cabrit07}.  It is still not clear what fraction of the mass
outflow rates observed on large-scales may be due to flows that are
magnetically connected to the star (as required in the present work),
as opposed to flows from the accretion disk
\citep[e.g.,][]{ferreira3ea06, kurosawa3ea06, kwan3ea07,
  kurosawa3ea11}.  Recent models of accretion-powered stellar winds by
\cite{Cranmer:2008p1657, Cranmer:2009p1647} exhibit outflows from the star itself with
$\chi\sim0.01$.  Thus, in the present work, we consider cases with
both $\chi=0.1$ and 0.01, to explore a possible range of this value.



     \subsection{Stellar Structure and Evolution} \label{sec_rstar}

The evolution of the stellar structure follows simple Kelvin-Helmholz
contraction of a polytrope (with index 3/2).  This treatment results
in an evolution of stellar radius that follows
\begin{equation}
\label{eq_rstar}
\frac{dR_{*}}{dt} = 2\frac{R_{*}}{M_{*}}\dot{M_{a}}(1-\chi) -
\frac{28\pi\sigma R_{*}^4 T_{e}^4}{3GM_{*}^2},
\end{equation}
where $R_*$ and $M_*$ are the stellar radius and mass, $T_e$ is the
effective temperature of the star, and $\sigma$ and $G$ are the
Stefan-Boltzmann constant and Newton's gravitational constant.  As in
Paper~I, we adopt an initial stellar radius of $8 R_\odot$ and
constant photospheric temperature $T_e = 4280$ K for all calculations,
so that the evolution of the star's structure resembles that from the
model of \citet{siess3ea00}, as the star descends along the Hayashi
track in the H-R diagram (see Fig.\ 3 in Paper~I).  The only
difference between this equation and the corresponding equation of
Paper~I is the addition of the factor $(1-\chi)$, which takes mass
loss into account.

     \subsection{Spin Evolution} \label{sec_spin}

The evolution of the angular spin rate of the star follows
\begin{eqnarray}
\label{eq_angmom}
\frac{d\Omega_{*}}{dt} = \frac {T_{*}}{I_*} - 
\Omega_{*} \left( \frac{\dot M_a}{M_*}(1-\chi) + 
\frac{2}{R_*} \frac{dR_*}{dt} \right), 
\end{eqnarray}
where $\Omega_*$ is the (solid body) angular rotation rate of the
star, $T_*$ is the net torque on the star, and $I_*$ is the stellar
moment of inertia, $I_*\equiv~k^2M_*R_*^2$, where $k$ is the
normalized radius of gyration (we adopt $k^2$ = 0.2).  The only
difference between this equation and the corresponding equation of
Paper~I is the addition of the factor $(1-\chi)$, which takes mass
loss into account.

It is often instructive to express the spin rate as a fraction of the
breakup speed, defined as the Keplerian velocity at the star's
equator.  This normalized spin rate is defined
\begin{eqnarray}
\label{eq_f}
f \equiv \Omega_* \sqrt{\frac{R_*^3}{G M_*}}.
\end{eqnarray}
Following Paper~I, we consider two cases with different
initial spin rates.  The two cases have initial fractional spin rates
of $f_0 = 0.3$ and 0.06, representing the two extremes of rapid and
slow initial rotation.

     \subsection{Torques on the Star} \label{sec_torque}

The present work considers the simultaneous effects of torques arising
in the magnetic star-disk interaction and angular momentum loss from
stellar winds.  In both cases, the angular momentum gain and loss of
the star is primarily transmitted by a magnetic field.  The model
assumes a rotation-axis-aligned dipole magnetic field, with a strength
of $B_*$ at the surface and equator of the star.  We consider two
different values of the magnetic field strength, $B_*=500$~G and
2000~G, in addition to a case with $B_*=0$ used for comparison.  The
model assumes that $B_*$ is constant in time for all cases.

           \subsubsection{Magnetic Star-Disk Interaction Torques} 
           \label{sub_sdint}

The torques arising from the magnetic interaction between the star and
accretion disk follow \citet{mattpudritz05}, and the governing
equations are given in Paper~I.  It is convenient to define the
corotation radius
\begin{eqnarray}
\label{eq_rco}
R_{\rm co} \equiv \left(\frac{G M_*}{\Omega_*^{2}}\right)^{1/3} = f^{-2/3} R_*,
\end{eqnarray}
which is the singular radius in the Keplerian disk that rotates at the
same angular speed as the star.  This location is physically
meaningful in the magnetic star-disk interaction because it is where
the differential rotation between the star and disk equals zero.

The magnetic field of the star is typically strong enough to be able
to truncate the disk at a distance of a few stellar radii.  The
truncation radius is denoted $R_t$.  The torque associated with disk
truncation and infall of material from $R_t$ to the stellar surface is
the accretion torque $T_a$.  There is also a magnetic torque
associated with the magnetic connection to a range of radii in the
disk.  This magnetic torque is denoted $T_m$.  The method and
equations used for calculating $R_t$, $T_a$, and $T_m$ are given in
Paper~I and are not modified for the present work.

Two key parameters involved in the calculation of the star-disk
interaction torques capture the physics of the magnetic coupling
strength (parameter $\beta$) and the opening of magnetic field lines
due to the differential rotation (parameter $\gamma_c$; see Paper~I
and \citealp{mattpudritz05}).  Parameter $\gamma_c$ reprents the
maximum ratio between the azimuthal and vertical magnetic field
components ($B_\phi/B_z$) threading the disk, in order for the dipolar
magnetic field lines to remain closed.  In the present work, we adopt
$\gamma_c = 1$ which corresponds to realistic expecations
\citep{uzdensky3ea02}.  Parameter $\beta$ is equivalent to the inverse
of the magnetic Reynolds number in the inner disk region.  Here we
adopt $\beta=0.01$, presented in Paper~I \citep[and][]{mattpudritz05}
as a reasonable guess for T Tauri systems.  The effect of varying
these two parameters was the subject of Paper~I.

          \subsubsection{Stellar Wind Torque} \label{sec_tw}

\begin{figure*}
\epsscale{\psiztwo}
\plotone{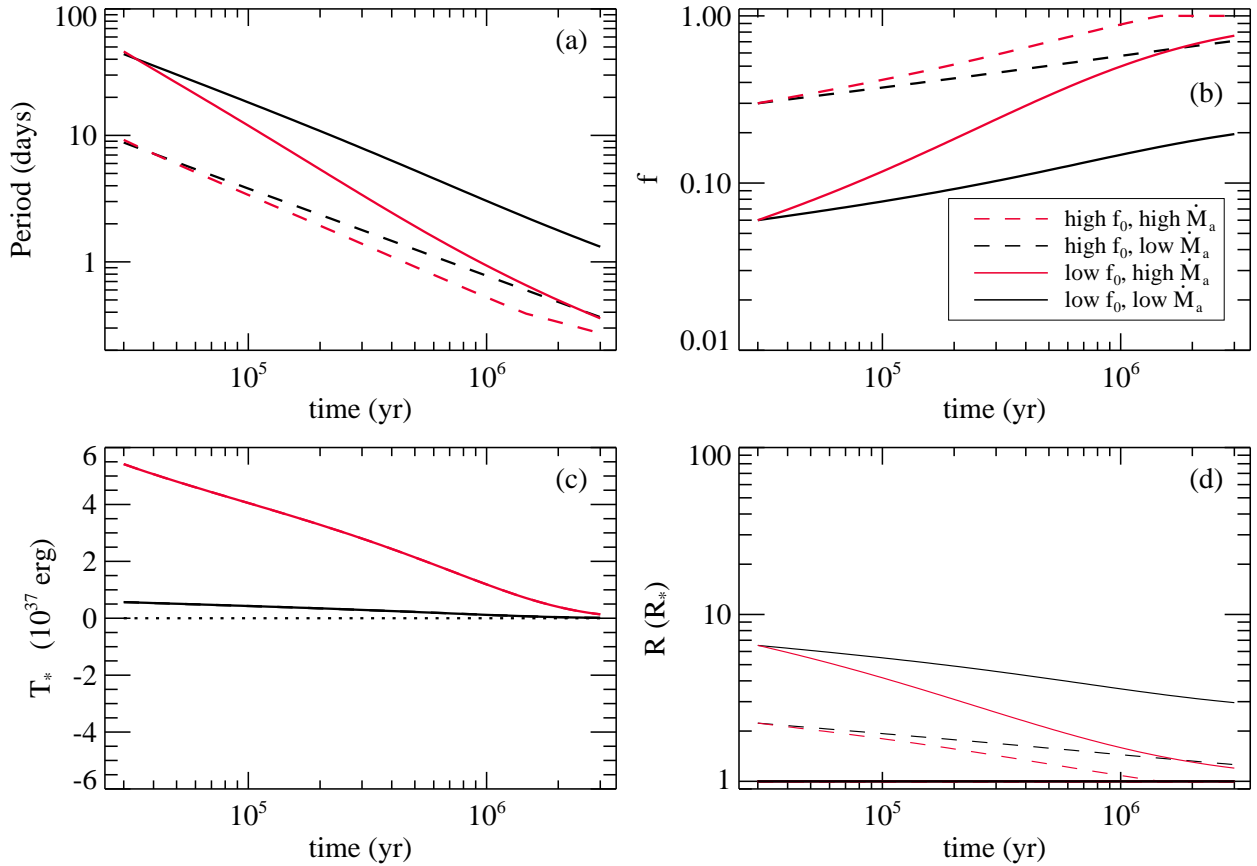}
\caption{Time-evolution of various quantities for the $B_*=0$ case.
 Panel (a) shows the rotation period in days, and panel (b) shows the
 spin rate expressed as a fraction of breakup speed.  The four lines
 shown represent the two choices of accretion history and two initial
 spin rates.  The red and black lines correspond to the high and low
 accretion rates, respectively.  The solid and dashed lines
 correspond to models with an initial spin rate of 0.06 and 0.3 times
 the breakup rate, respectively.  Panel (c) shows the net torque on
 the star (eq.\ [\ref{eq_torque}]) for each case.  In this panel, the
 dashed and solid lines of a given color lie on top of each other,
 since the spin rate does not affect the torque, when $B_*=0$.  The
 horizontal dotted line indicates $T_*=0$ for clarity.  Panel (d)
 shows the location of the truncation radius ($R_t$, thick lines) and
 the corotation radius ($R_{\rm co}$, thin lines; eq.\
 [\ref{eq_rco}]), in units of the stellar radius.  Here, $R_t=R_*$
 for all models, since there is no magnetic field to truncate the
 disk above the stellar surface, so all thick lines (solid and
 dashed, black and red) overlap.}
\label{fig_b0}
\end{figure*}

\begin{figure*}
\epsscale{\psiztwo}
\plotone{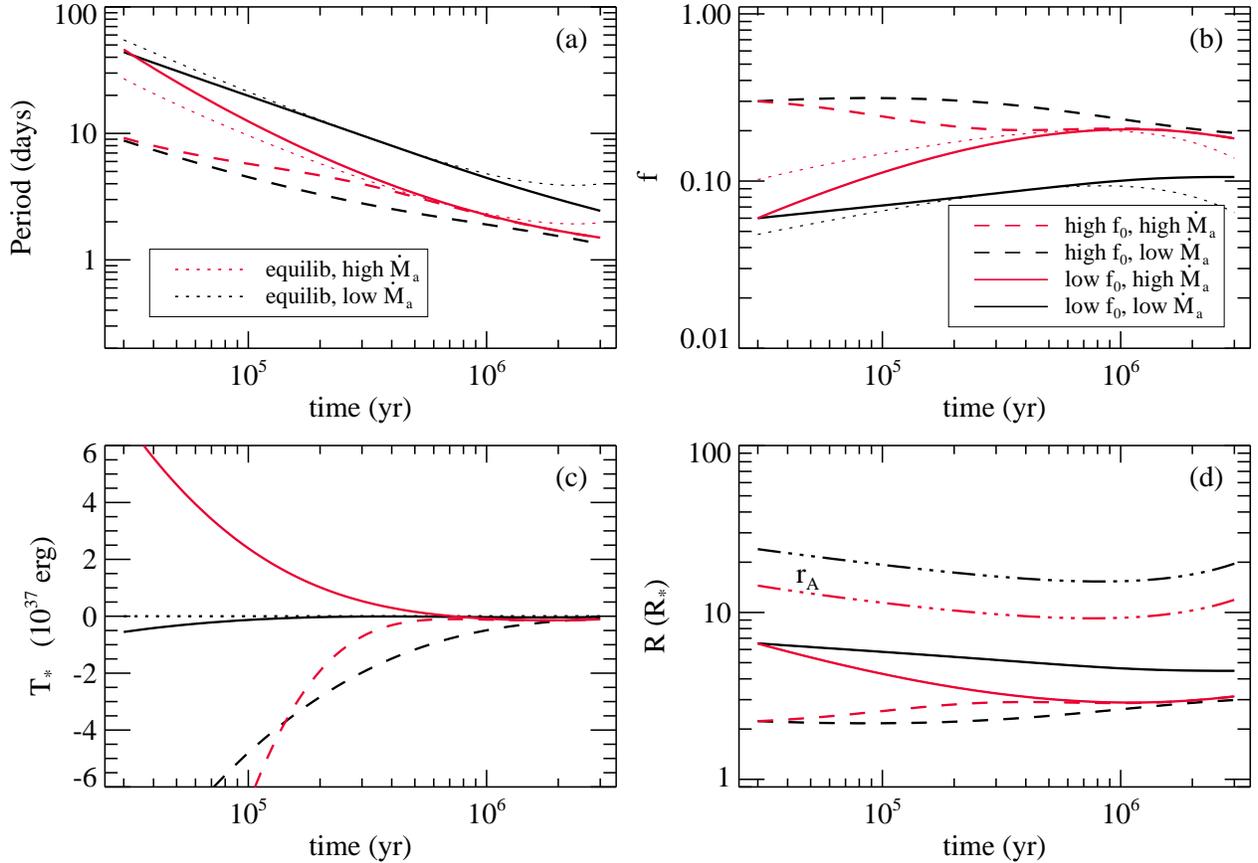}
\caption{Results for $B_*=500$~G and $\dot M_w / \dot M_a = 0.1$, in
the same format as Figure \ref{fig_b0}.  In panels (a) and (b), the
dotted red and black lines show the theoretical equilibrium rotation
rates (eq.\ [\ref{eq_feq}]) for the high and low accretion rates,
respectively.  In panel (d), the dash-triple-dotted red and black
lines show the stellar wind Alfv\'en radius (\S \ref{sec_tw}) for
the high and low accretion rates, respectively.  Also in panel (d),
the thin and thick solid lines of a given color and type are
indistinguishable, since $R_t$ is very close to $R_{\rm co}$ in all
four cases.}
\label{fig_10b500}
\end{figure*}

\begin{figure*}
\epsscale{\psiztwo}
\plotone{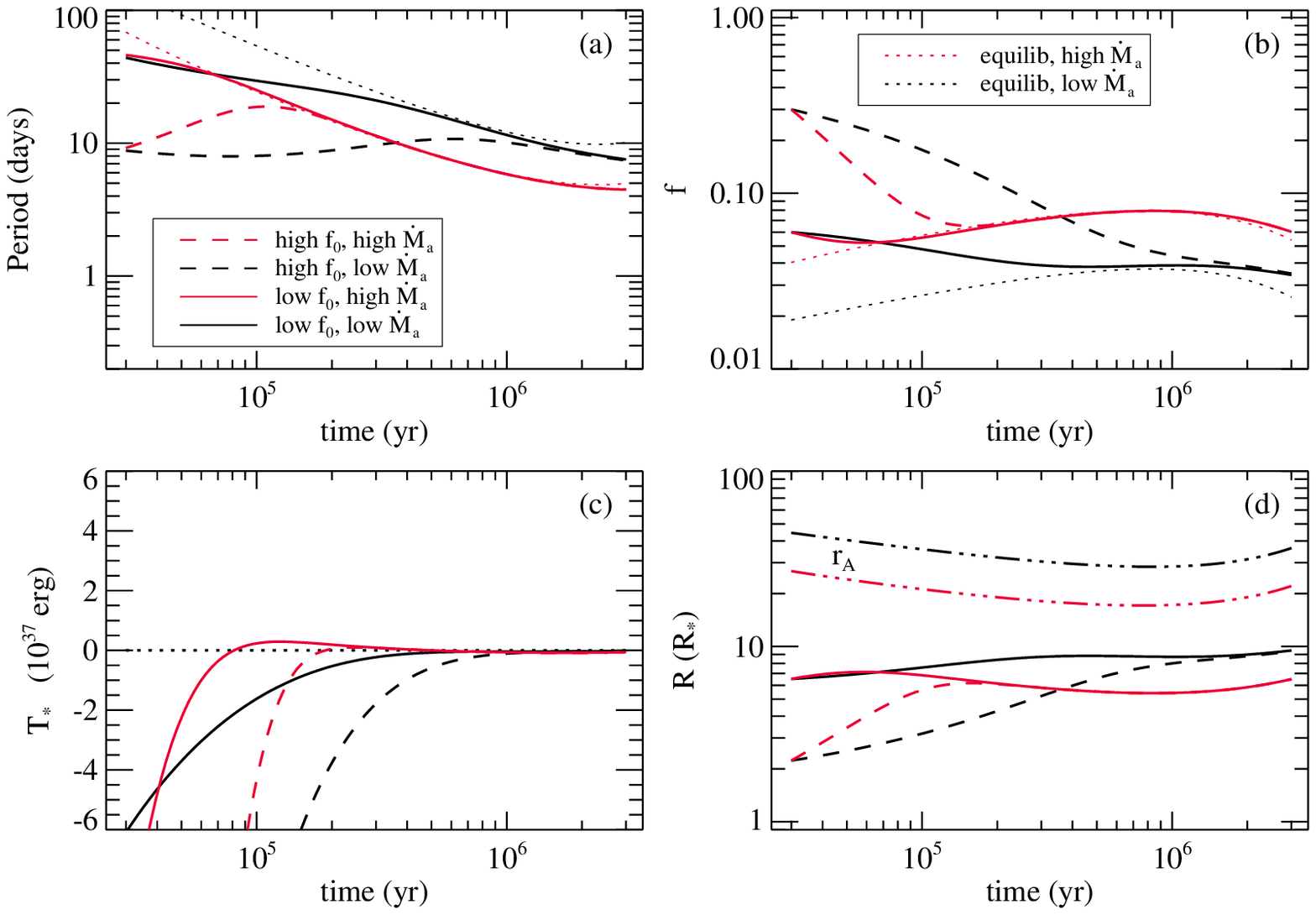}
\caption{Results for $B_*=2000$~G and $\dot M_w / \dot M_a = 0.1$, in
the same format as Figure \ref{fig_10b500}.  In panel (d), the thin
and thick lines of a given color and type are indistinguishable,
since $R_t$ is very close to $R_{\rm co}$ in all four cases.}
\label{fig_10b2000}
\end{figure*}

\begin{figure*}
\epsscale{\psiztwo}
\plotone{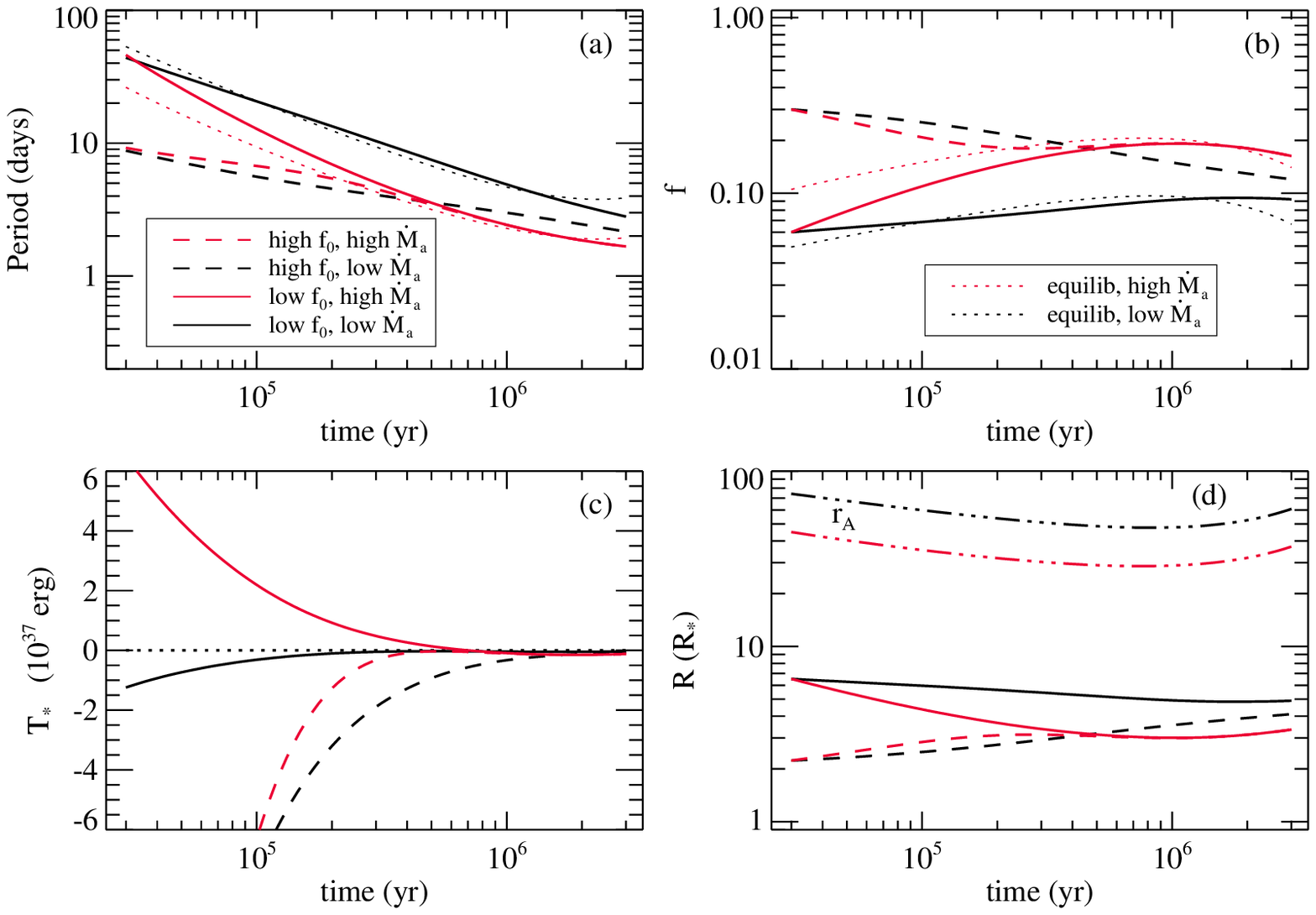}
\caption{Results for $B_*=2000$~G and $\dot M_w / \dot M_a = 0.01$, in
the same format as Figure \ref{fig_10b500}.  In panel (d), the thin
and thick lines of a given color and type are indistinguishable,
since $R_t$ is very close to $R_{\rm co}$ in all four cases.}
\label{fig_01b2000}
\end{figure*}

In order to calculate the angular momentum loss via stellar winds, we
use the stellar wind torque formulation of \citet{mattpudritz08II}.
This formulation is based upon analytic calculations going back to
\citet{weberdavis67}, in which the torque from a one-dimensional wind
is given by
\begin{eqnarray}
\label{eq_tw}
T_w = - \dot M_w \Omega_* r_{\rm A}^2,
\end{eqnarray}
where $r_{\rm A}$ is the Alfv\'en radius, the location where the wind
speed equals that of magnetic Alfv\'en waves.  In real stellar winds,
the flow is aspherical, due to (e.g.) rotation and magnetic geometry.
\citet{mattpudritz08II} computed solutions for two-dimensional
(axisymmetric) solar-like stellar winds, using numerical
magnetohydrodynamic simulations.  They used the simulations to compute
$T_w$ and $\dot M_w$ and then used equation (\ref{eq_tw}) to define
$r_{\rm A}$, which in this context represents the mass-loss-weighted
average of the Alfv\'en radius in the multi-dimensional flow.  For
variations in the magnetic field strength, stellar radius, surface
gravity, and mass loss rate, the solutions for $r_{\rm A}$ are
well-fit by the following
\begin{eqnarray}
\label{eq_ra}
{r_{\rm A} \over R_*} = K \left({ {B_*^2 R_*^2} \over 
                               {\dot M_w v_{\rm esc}} }\right)^m,
\end{eqnarray}
where $K$ and $m$ are fit constants.  For the cases with a dipole
magnetic field, the simulations of \citet{mattpudritz08II} indicate $K
\approx 2.11$ and $m \approx 0.223$, which we adopt in this work.  A
similar numerical study of the torque from the winds of massive stars
\citep{uddoula3ea08, uddoula3ea09} found a relationship similar to
equation (\ref{eq_ra}), with $m=1/4$ and $K$ of the order of unity.
By contrast, studies of cold, magnetocentrifugally-driven flows
\citep[e.g.,][]{pelletierpudritz92,Spruit:1996p3813,andersonea05}
imply $m=1/3$.

          \subsubsection{Total Torque and Equilibrium Spin Rate} \label{sub_equilib}

The total torque on the star, used in equation (\ref{eq_angmom}), is
the sum of the star-disk interaction torques and the stellar wind
torque,
\begin{eqnarray}
\label{eq_torque}
T_* = T_a + T_m + T_w.
\end{eqnarray}

The spin rate of any star-disk interacting system generally evolves in
such a way as to tend toward a theoretical ``equilibrium spin rate,''
at which the net torque on the star is zero:
\begin{eqnarray}
\label{eq_t0}
T_* = 0 \;\;\;\;\;\; {\rm(in \; spin \; equilibrium)}.
\end{eqnarray}
For the values of the magnetic field opening and coupling parameters
considered in this paper ($\gamma_c=1$ and $\beta=0.01$),
\citet{mattpudritz08III} showed that the magnetic torque $T_m$ is
generally negligible compared to the stellar wind torque $T_w$.  Thus,
the spin equilibrium is approximately characterized by $T_a=-T_w$.
This predicts an equilibrium spin rate, expressed as a fraction of
breakup speed, given by (eq.\ [18] of \citealp{mattpudritz08III}),
\begin{eqnarray}
\label{eq_feq}
%
f_{\rm eq, sw} = K^{-3/2} \chi^{(6m-3)/4} \Psi^{-3m/2},
\end{eqnarray}
where
\begin{eqnarray}
\label{eq_psi}
\Psi \equiv { {B_*^2 R_*^2} \over {\dot M_a v_{\rm esc}}}.
\end{eqnarray}
Equation (\ref{eq_feq}) assumes that the truncation radius is always
near the corotation radius, which turns out to be true in the models
with non-zero magnetic field presented in this paper.  Below, we will
use the predicted equilibrium spin rate of equation (\ref{eq_feq}) to
compare with the results of our spin evolution models.

    \subsection{Numerical Method}

    The coupled equations (\ref{eq_mdotstar}), (\ref{eq_rstar}), and
    (\ref{eq_angmom}) describe the evolution of the system in time. We
    wrote a computational code that solves these simultaneously, using
    the fourth-order Runge-Kutta scheme of \citet{pressea94}, starting
    from $t_0=3\times10^4$ yr and ending at 3 Myr.  The code is
    described in more detail in Paper~I.

\section{Results} \label{sec_results}

Table \ref{tab_parms} contains the parameters for each case presented
in this section, listed in order of their presentation and grouped by
the figures in which the results appear.  For each case, we present
the evolution of the system for the 4 possible combinations of 2
different initial mass accretion rates (parameterized by $\dot M_{a0}$) and 2
different initial spin rates ($f_0$).  The ``$B_*=0$'' case is
identical to that of Paper~I, and we include it here to facilitate
comparison with the new results (cases W1--W3).

\begin{deluxetable}{ccccccc}
\tablewidth{0pt}
\tablecaption{Model Parameters \label{tab_parms}}
\tablehead{
\colhead{Case} &
\colhead{$\chi$} &
\colhead{$B_*$ (Gauss)} &
\colhead{$\dot M_{a0} (M_{\odot}$/yr)} &
\colhead{$f_0$} &
\colhead{Figure} 
}

\startdata

$B_*=0$ & 0 & 0 & $10^{-8}$, $10^{-7}$ & 0.06, 0.3 & \ref{fig_b0} \\
W1   & 0.1       & 500      &$10^{-8}$, $10^{-7}$ & 0.06, 0.3 & \ref{fig_10b500} \\
W2   & 0.1       & 2000     & $10^{-8}$, $10^{-7}$ & 0.06, 0.3 & \ref{fig_10b2000} \\
W3   & 0.01       & 2000     & $10^{-8}$, $10^{-7}$ & 0.06, 0.3 & \ref{fig_01b2000}
\enddata


\end{deluxetable}

It is instructive first to examine the non-magnetic case, $B_*=0$, in
which there are no magnetic fields.  Figure \ref{fig_b0} illustrates
the evolution of the spin rate---expressed as the spin period (panel
(a)) and as a fraction of breakup speed (panel (b))---the total torque
experienced by the star (panel (c)), and the radial locations of the
disk inner edge ($R_t$) and corotation radius ($R_{\rm co}$) (panel
(d)).  In the $B_*=0$ case, the disk extends all the way to the surface
of the star ($R_t=R_*$, as indicated by the thick lines in panel d),
and the torque experienced by the star (panel c) is solely the
accretion torque, $T_a=\dot M_a\sqrt{GM_*R_t}$ (since $T_m=T_w=0$).

In all 4 models shown in Figure \ref{fig_b0}, the stellar spin rate
increases monotonically in time, due to the accretion of disk material
with high specific angular momentum and due to the contraction of the
star.  Only the most extreme model (with the slowest initial spin and
lowest accretion rate, black solid line in panels (a) and (b)) has a spin
period in the range 1--3 days during the age range of 1--3 Myr.  The
other 3 models have spin periods substantially less than a day, in the
same age range.

All subsequent cases (W1--W3) have a non-zero magnetic field and use
$\gamma_c=1$ and $\beta=0.01$ for the star-disk magnetic connection
parameters, which are appropriate values to include the effects of
strong magnetic coupling to the disk and the opening of magnetic field
lines by azimuthal twisting \citep[see][and Paper~I]{mattpudritz05}.
Thus, in addition to the accretion torque, the model stars experience
spin-up and spin-down torques ($T_m$) associated with a magnetic
connection over a finite region of the disk.  Also, cases W1--W3
include the mass loss ($\chi$) and torque ($T_w$) due to stellar winds
(as described in \S \ref{sec_model}).

Figure \ref{fig_10b500} shows the results for the four models in the
W1 case, in which the star has a 500 Gauss magnetic field and stellar
wind mass loss rate that is 10\% of the disk accretion rate
($\chi=0.1$).  In this Figure, the line colors and styles have the
same meaning as Figure \ref{fig_b0}, with a few additions.  Panels (a)
and (b) include dotted lines that indicate the equilibrium spin value
predicted by equation (\ref{eq_feq}) for the high and low mass
accretion rate models (red and black lines, respectively).  In panel
(d), the dash-triple-dotted lines indicate the location of the
Alfv\'en radius in the stellar wind (eq.\ [\ref{eq_ra}]) for the
models with high and low accretion rate (and thus high and low wind
outflow rate; red and black lines, respectively).

By comparing Figures \ref{fig_b0} and \ref{fig_10b500}, it is clear
that the stellar spin evolution is substantially influenced by the
presence of a 500 G field and accretion-powered stellar wind with
$\chi=0.1$.  In the age range of 1--3 Myr, all four models in Figure
\ref{fig_10b500} have spin periods in the range of 1-5 days, which
corresponds to spin rates between 10\% and 20\% of breakup.  As
visible in panel (d), the truncation radius is very close to the
corotation radius (such that the thick and thin lines overlap in the
plot), for all models at all times.  The Alfv\'en radius in the
stellar wind (panel (d)) for all models remains in the range of
approximately 10-30 $R_*$.

Panels (a) and (b) show the equilibrium spin rate (dotted lines)
predicted by equation (\ref{eq_feq}) under the assumption that the
spin evolution is dominated by a balance between the accretion
(spin-up) torque and stellar wind (spin-down) torque.  In Figure
\ref{fig_10b500}, the two models corresponding to a high accretion
rate (red solid and dashed lines) approach the equilibrium spin rate
at an age of a few times $10^5$ yr.  After this time, the initial
condition for the spin rate of these models has effectively been
``erased,'' in a sense that the subsequent spin rate is insensitive to
the initial rate.  For the models with low accretion rate (black solid
and dashed lines), the torque is not strong enough to drive the spin
rate to the equilibrium value within 3 Myr.  The case with a slow
initial spin rate (black solid line) evolves near the equilibrium
value, but this is only due to a coincidence between the initial and
equilibrium spin rates.

Figure \ref{fig_10b2000} shows results for the W2 case, which is the
same as W1, except that the magnetic field is 4 times stronger
($B_*=2000$ G).  By comparing Figures \ref{fig_10b500} and
\ref{fig_10b2000}, it is clear that the stronger magnetic field results
in slower spin rates.  In the age range of 1--3 Myr, all four models
in Figure \ref{fig_10b2000} have spin periods in the range of 4--12
days, corresponding to spin rates between 3\% and 8\% of breakup.  The
models with high accretion rate in the W2 case
(Fig.\ \ref{fig_10b2000}, red solid and dashed lines) approach
the equilibrium spin rate after $\sim 10^5$ yr, sooner than the weaker
field case (W1).  Furthermore, the models with low accretion rate
(black solid and dashed lines) approach the equilibrium spin rate
after $\sim 10^6$ yr.  In all models and at all times, the truncation
radius is very close to the corotation radius (such that the thick and
thin lines overlap in panel (d)), and the Alfv\'en radius in the winds
remains in the range of 15--50 $R_*$.

Figure \ref{fig_01b2000} shows results for the W3 case, which is the
same as W2, except that the stellar wind outflow rate is 10 times less
($\chi=0.01$).  A lower wind mass outflow rate, if all else is equal,
means a lower spin-down torque from the stellar wind.  Thus, the
accretion-powered stellar wind in the W3 case is less effective at
spinning down the star than the W2 case.  A comparison between Figures
\ref{fig_01b2000} and \ref{fig_10b500} reveals that the W3 case is
qualitatively and quantitatively very similar to the W1 case, which
has a higher wind outflow rate but a weaker magnetic field.  The
similarity in the spin evolution of these two cases is a consequence
of the dependence of the stellar wind torque on the parameters
changed.  Specifically, equations (\ref{eq_tw}) and (\ref{eq_ra}) and
$m=0.223$ results in $T_w \propto \dot M_w^{0.55} B_*^{0.89}$,
which means that the factor of 10 difference in $\dot M_w$ is almost
completely compensated by the factor of 4 difference in $B_*$.  While
the stellar wind torques are similar in these two cases, the Alfv\'en
radius in the wind has a different dependence on these parameters, and
thus the models in the W3 case have a much larger Alfv\'en radii than
the W1 case (compare the dash-triple-dotted lines in panel (d)).

It is instructive to compare the cases presented here with the cases
in Paper~I.  Case W1 is identical to case O1 in Paper~I, except that
W1 includes the effect of an accretion-powered stellar wind, via a
nonzero value of the mass-loss parameter $\chi$.  Likewise, cases W2
and W3 are similar to O2 of Paper~I.  A comparison between W1--W3
(Figs.\ \ref{fig_10b500}--\ref{fig_01b2000}) and O1/O2 (Figs. 6 and 7
of Paper~I) indicates that the cases with stellar winds spin
substantially slower than the cases without.  For the range of
magnetic field strengths, accretion rates, initial spin rates, and
mass outflow rates considered in cases W1--W3, the models exhibit
rotation periods within the range of approximately 1--10 days in the
age range of 1--3 Myr.  These rotation periods correspond to spin
rates in the range of approximately 5\%--20\% of breakup speed and lie
within the bulk of the distribution observed in stars of similar age
and mass (see \S \ref{sec_intro}).

For all of the models in cases W1--W3 that approach an equilibrium,
the spin rates closely follow the prediction of equation
(\ref{eq_feq}).  Since that equation assumes a balance solely between
the accretion torque and a stellar wind torque ($T_*=-T_w$), the fact
that some models follow this prediction indicates that the magnetic
torque from the star-disk connection ($T_m$) is negligible in those
models.  More generally, \citet{mattpudritz08III} showed that, for the
strong coupling case considered here (i.e., $\beta=0.01$ and
$\gamma_c=1$), the stellar wind torque dominates over the spin-down
portion of the magnetic torque $T_m$ as long as the Alfv\'en radius in
the stellar wind ($r_{\rm A}$) is $ \la 84 R_*$.  This is true during
the entire evolution for all models presented here (see
dash-triple-dotted lines in panel (d) of
Figs.\ \ref{fig_10b500}--\ref{fig_01b2000}).  Thus, it is the stellar
wind torque, not the star-disk magnetic connection, that is
responsible for the slower spin of these models relative to the
non-magnetic case.

\vspace{1.5cm}

\section{Comparison Between APSW and Disk Locking Models} \label{sec_comparison}

For the magnetic torque model and chosen parameters presented in
sections \ref{sec_model} and \ref{sec_results}, we showed that the
spin-down torque from APSWs completely dominates over any spin-down
torques that may arise from the magnetic connection between the star
and disk.  Thus, the evolution of the stellar spin is primarily
determined by a competition between the spin-down torques from an APSW
and a spin-up via the accretion torque (hereafter, the ``APSW
model'').  However, given the large number of unknowns regarding real
systems, it is instructive to compare the predictions of different
star-disk interaction models for a range of possible conditions and
assumptions.

As discussed earlier, in these systems the spin rate of the star
always tends toward the equilibrium spin rate at which the net
torque on the star would be zero.  If the spin-up or spin-down time is
short (i.e., for a strong net torque) compared to the evolution
timescale, the star's spin will remain close to this equilibrium
value.  Whether or not the torque is strong enough to keep the star
spinning near equilibrium depends upon many factors (e.g., the
accretion rate $\dot M_a$ and magnetic field strength $B_*$), which
also may be varying in time.  Thus it is not generally expected that
all systems should be in equilibrium.  However, for the purpose of
comparing different models, it is convenient and particularly
instructive to compare the predictions of the equilibrium spin rate
for each model.  When a particular model predicts a slower spin rate
compared to other models, it simply means that the spin-down torques
are generally stronger in that model.

\begin{figure}
\epsscale{1.2}
\plotone{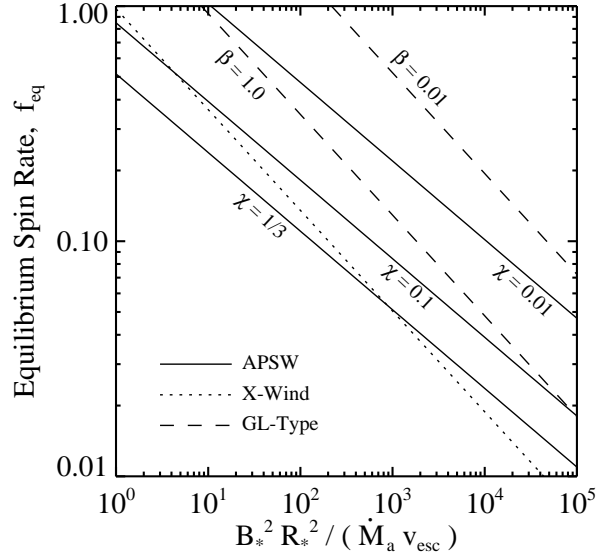}
\caption{Predicted equilibrium spin rates of different star-disk
  interaction models as a function of the dimensionless parameter
  $\Psi$ (eq.\ [\ref{eq_psi}]).  The solid lines show the prediction
  of the APSW model (determined by eq.\ [\ref{eq_feq}], with $K=2.11$
  and $m=0.223$) for three different values of $\chi$, as indicated.
  The dashed lines show the predictions for the Ghosh \& Lamb-type
  disk locking model of \citet{mattpudritz05} (determined by
  eq.\ [\ref{eq_feqdl}], with $C=C(\beta, \gamma_c)$) for $\gamma_c=1$
  and two different values of the magnetic coupling parameter $\beta$,
  as indicated.  The dotted line shows the prediction of the X-wind
  model (determined by eq.\ [\ref{eq_feqdl}] with $C=0.974$).}
\label{fig_feqs}
\end{figure}

The equilibrium spin rate predicted by the APSW model, expressed as a
fraction of breakup speed, is given in equation (\ref{eq_feq}).  In
Figure \ref{fig_feqs}, the solid lines show this predicted equilibrium
spin rate as a function of the dimensionless parameter $\Psi$ and for
3 different values of $\dot M_w / \dot M_a \equiv \chi =$ 1/3, 0.1,
and 0.01.  As discussed in section \ref{sec_model}, we adopt values of
$K=2.11$ and $m=0.223$, which results in a power-law slope of $-3m/2
\approx -0.33$ in the Figure.  The plot covers a range in $\Psi$ from
1 to $10^5$, which is chosen to approximately reflect the possible
range in real systems, obtained by choosing extreme values of
parameters from the range that is observed for T Tauri stars
\citep[e.g.,][]{hartmannea98, gullbringea98, johnskrullgafford02,
  robbertoea04, siciliaaguilarea06, Bouvier:2007p3031}.

Of the various star-disk interaction studies in the literature, there
are two other types of general models that predict an equilibrium spin
rate, and with which we will compare the predictions of the APSW
model.  These are the Ghosh \& Lamb type (GL-type) models \citep[e.g.,
Paper~I; ][]{camenzind90, konigl91, cameroncampbell93, cameron3ea95,
  yi94, yi95, lovelace3ea95, armitageclarke96, mattpudritz05,
  zweibel3ea06}, which are based on the original accreting magnetized
star model of \citet{ghoshlamb78}, and the X-wind model
\citep{shuea94, shuea94II, najitashu94, ostrikershu95, shuea95,
  mohantyshu08}.  Both the GL-type and X-wind models assume that the
magnetic connection between the star and disk alone is responsible for
removing the necessary angular momentum, and both neglect the
influence of stellar winds.  In their equilibrium state, these are
often referred to as ``disk locking'' models \citep{choiherbst96}.
The predicted equilibrium spin rate for both types of models can be
expressed as
\begin{eqnarray}
\label{eq_feqdl}
%
f_{\rm eq, dl} = C \Psi^{-3/7},
\end{eqnarray}
where $C$ is a dimensionless constant that takes into account various
assumptions in the models, and $\Psi$ is the same as defined in
equation (\ref{eq_psi}).

The various disk-locking models in the literature differ in their
assumptions, which is reflected in a different value of $C$.  The
star-disk torque model of \citet{mattpudritz05} (used in the present
paper) is a GL-type model that includes the effect of magnetic field
opening (via the parameter $\gamma_c$) and magnetic coupling strength
(via the parameter $\beta$).  That work showed that the
constant\footnote{For convenience here, we have absorbed all constant
  factors into the value of $C$ in equation (\ref{eq_feqdl}), as
  compared to equation (25) of \citet{mattpudritz05}.}  $C$ is a
function of only $\gamma_c$ and $\beta$, and different choices of
these can result in $C$ values that span the range of GL-type models
presented in the literature.  \citet{uzdensky3ea02} demonstrated that
a value $\gamma_c = 1$ (adoped in the present work) is the most
appropriate to properly take into account the effect of the opening of
magnetic field lines due to differential twisting between the star and
disk.


The dashed lines in Figure \ref{fig_feqs} show the prediction of
GL-type models for $\gamma_c=1$ and two different values of $\beta$
(0.01 and 1.0, corresponding to $C \approx 10.1$ and $C \approx
2.49$, respectively).  The upper-most line corresponds to the GL-type
model used to compute the magnetic star-disk torques in the
spin-calculations of the present paper (i.e., for $\beta = 0.01$).  It
is clear that this GL-type model predicts an equilibrium spin rate
that is faster than the APSW model, for the entire plotted range of
$\Psi$, and as long as $\chi$ is greater than a few thousanths.  This
is just another way of demonstrating that, when the mass loss is high
enough, APSWs are much more effective at spinning down stars than a
GL-type star-disk interaction alone.

The effective value of $\beta$ in real systems is not well
constrained.  In the GL-type models with $\gamma_c=1$,
\citet{mattpudritz05} showed that the value of $\beta=1$ results in
the minimum value for $C$ and thus represents the case where the
GL-type model torques would be the most effective at spinning the star
down.  The bottom-most dashed line in Figure \ref{fig_feqs} shows the
prediction for this special case.  It is clear that the APSW model
generally predicts a slower spin rate for $\chi \sim 0.1$ than the
GL-type models with any value\footnote{This statement is true for
  $\gamma_c=1$.  Most of the GL-type models in the literature have not
  taken into account the opening of the magnetic field, which is
  equivalent to assuming $\gamma_c=\infty$.  In this case, and
  especially when $\beta$ is taken to be small \citep[e.g., as in the
  model of][]{cameroncampbell93}, the spin-down torque is
  (artificially) enhanced and can be arbitrarily large.} of $\beta$.
For $\chi \sim 0.01$, the APSW model predictions are of a similar
magnitude as those of a GL-type model with $\beta=0.1$.  While the
effective value of $\beta$ in real systems is uncertain, it seems
likely that the inner disks of young stars exist in a state of high
magnetic Reynolds number ($\beta \ll 1$), and thus that the GL-type
spin-down torques are relatively weak.

In the X-wind model, the angular momenum is assumed not to accrete
onto the star, but rather is intercepted at the inner edge of the disk
(the ``X-point'') and then carried out of the system by a wind (the
``X-wind'') from the X-point.  The fiducial model of \citet{shuea94}
and \citet{ostrikershu95} supposes that the wind outflow rate from the
X-point is $\sim 1/3$ times the accretion rate, which results in a
predicted equilibrium spin rate that is given by equation
(\ref{eq_feqdl}), with a value of $C\approx0.974$.  The dotted line in
Figure \ref{fig_feqs} shows this predicted spin rate of the X-wind
model.  It is clear that this model also predicts slower spin rates
than the GL-type models.

Both the APSW and X-wind models employ a wind to remove angular
momentum from the system.  Thus, it is perhaps not surprising that the
predicted spin rate of the APSW model with the same mass outflow rate
as the X-wind, $\chi=1/3$ (bottom solid line in Fig.\ \ref{fig_feqs}),
is quantitatively similar to the predictions of the X-wind.
Furthermore, it can be shown that the factor $C$ in the X-wind model
depends upon the mass loss rate to the power of $-3/7$, which is very
close to the dependence of the APSW model prediction on the mass loss
rate (in eq.\ [\ref{eq_feq}], $f\propto\chi^{-0.42}$, for the adopted
value of $m\approx0.223$).

Although the predicted equilibrium spin rates are quantitatively
similar, the physical picture of APSW and X-wind models have some
important differences.  In the X-wind, the wind is magnetically
connected to, and removes angular momentum from, the X-point (not from
the star, as in a stellar wind).  The X-wind model is developed under
the assumption that there is no net torque on the star and that the
X-point is located at the corotation radius.  Thus, the balancing of
angular momentum is evaluated at the X-point (rather than on the
star), and this leads to a requirement for a certain amount of stellar
magnetic flux to be pinched, or ``trapped,'' within the X-point.
Having the flux is trapped at the X-point avoids the large-scale
opening of magnetic field (which is a problem for the GL-type models)
and also produces a geometry in which magneto-centrifugal forces may
provide a natural explanation for the acceleration of the wind.
However, the biggest problem for the X-wind model is that the amount
of flux trapping required at the X-point raises questions of dynamical
stability \citep[e.g.,][]{uzdensky04}, and such a magnetic field
configuration has not yet been demonstrated to exist in a dynamical
model.  Furthermore, since the model assumes no net torque on the star
(i.e., it assumes the star to be in spin equilibrium), the model
provides little information about how angular momentum may flow from
accreting material outward to the X-point, and accross the X-point, in
order to be removed in the X-wind and/or via transport processes in
the disk \citep[][]{Ferreira:2000p1706}.  The assumption of
equilibrium also means that the model cannot address how these systems
evolve (e.g., one cannot make plots such as panels [c] in Figs.\
\ref{fig_b0}--\ref{fig_01b2000}), nor whether or how a system will
reach or approximate the equilibrium state and magnetic field
configuration described.


By contrast, the APSW model calculates the spin-up and spin-down
torques acting on the star from various processes (i.e., at different
geographical locations on the star).  The theoretical equilibrium
state corresponds to when the net torque is zero, but the model can be
used to compute the net torque in any state.  The magnetic field
configuration is that of the stellar magnetic field, which has a
significant amount of flux that is opened by various dynamical
processes \citep[e.g., see discussion in][]{mattpudritz05}.
Numerical, dynamical simulations \citep[e.g.,][]{hayashi3ea96,
  goodson3ea97, millerstone97, goodson3ea99, fendtelstner99,
  fendtelstner00, mattea02, kuker3ea03, VonRekowski:2004p2874,
  VonRekowski:2006p2879, long3ea05, romanovaea05, bessolazea08,
  zanniferreira09, fendt09, romanovaea09, longea11, romanovaea11}
generally show that such a magnetic field configuration is a natural
and general consequence of the star-disk interaction.  It is the open
stellar field lines that are primarily responsible for the spin-down
torque on the star, via a wind flowing along the field.  The amount of
magnetic flux participating in the wind is not a free parameter, but
is determined in a dynamically self-consistent way via the MHD
simulations used to compute equation (\ref{eq_tw}) for the stellar
wind torque \citep[][]{mattpudritz08II}.  Thus, there is no conceptual
nor dynamical problem with the magnetic field configuration of the
APSW model.  However, the biggest open question for APSW model is
whether and/or how the accretion power in the system can drive a wind
that is magnetically connected to the star and that has a mass outflow
rate high enough to extract significant angular momentum.  This is
discussed further in section \ref{sec_conclusions}.

It is clear from Figure \ref{fig_feqs} that the APSW predicts a slope
that is somewhat shallower than the disk-locking models.  This fact
could in principle be used to discriminate between the models
\citep[e.g., similar to][]{johnskrullgafford02}, although this may not
be possible with existing data.  The difficulty is due to at least 3
factors: a) there are large observational uncertainties for the
quantities plotted in Figure \ref{fig_feqs} and few stars for which
these quantities have been measured; b) not all systems are expected to
be in spin-equilibrium; and c) these systems are typically highly
time-variable \citep[e.g.,][]{Hartmann:1997p3222}, which means that a
realistic spin-equilibrium (net zero torque) state may only exist in a
time-averaged sense---or in a statistical sense, for a large
collection of stars.  Furthermore, the predicted slope for the APSW
model may be different, depending upon a few unknowns.  The form of
equation (\ref{eq_feq}) assumes that the disk truncation radius is
close to the corotation radius \citep{mattpudritz08III}, which will
not necessarily be true in all systems.  Also, the value of the power
law parameter $m$ depends upon the conditions in the stellar wind, and
a number of factors (discussed in \S \ref{sec_conclusions}) may affect
this value.

Finally, we note that in the interest of performing direct and
quantitative comparisons, this section has focused on the three
classes of star-disk interaction models for which a simple prediction
for a hypothetical equilibrium spin rate is possible.  Thus, there are
a number of other ideas in the literature that we have neglected and
that nonetheless deserve further attention.  For example,
time-variability of the accretion rate onto the star may strongly
influence stellar spin evolution \citep[e.g.,][]{popham96,
  DAngelo:2010p3127}, and the dynamics of the interface region that
exists between the open magnetic flux threading the star and disk may
also be important for extracting angular momentum \citep[e.g.,][Zanni
\& Ferreira, in preparation]{hiroseea97, Ferreira:2000p1706,
  romanovaea09}.


\section{Summary and Discussion} \label{sec_conclusions}

Paper~I presented a model for computing the evolution of mass, radius,
and spin rate of an accreting, one solar mass star during the Hayashi
phase (from $3\times10^4$ yr to $3$ Myr).  The spin of the star
evolves from its initial value due to changes in the moment of inertia
(due to stellar contraction) and to external torques.  To compute the
external torques, Paper~I included only the magnetic star-disk
interaction torque formulation of MP05, and neglected any torque along
open field lines.  That work showed that, for strong magnetic coupling
to the disk ($\gamma_c = 1$ and $\beta = 0.01$), cases with either
$B_*=500$G or 2000G result in spin evolution that is only slightly
changed from the non-magnetic case.  In other words, when the magnetic
coupling is strong and $B_* \la 2000$G, the spin-down torque arising
in the star-disk interaction is negligible for the spin rates and mass
accretion rates considered.

The goal of the present paper was to determine under what conditions
the additional spin-down torque from stellar winds may significantly
affect the spin evolution and produce a range of spin rates consistent
with the observed range.  Thus, we have extended the model of Paper~I
(as described in \S \ref{sec_model}) to include an additional
spin-down torque due to the stellar wind.  As stellar wind torques
depend upon most of the same parameters as star-disk interaction
torques, including a stellar wind has introduced only one new
parameter, the mass outflow rate of the wind.  Here, we consider the
case where the mass outflow rate is some fixed fraction ($\chi$) of the
accretion rate, as may be expected for a wind that is somehow driven
by accretion power \citep{mattpudritz05l}.

In section \ref{sec_results}, we found that the cases with $B=2000$G
and mass loss rates of $\chi=0.1$ resulted in spin rates in the range
of $\sim5$--10 days, in the age range of 1--3 Myr.  For the two cases
with either a lower mass loss rate ($\chi=0.01$) or magnetic field
strength ($B=500$G), the resulting spin rates were in the range of
1--5 days (for 1--3 Myr ages).  These two cases exhibited similar spin
evolution because the stellar wind torques were nearly identical in
both cases (i.e., there is a degeneracy between $B_*$ and $\dot M_w$
in the stellar wind torque, eq.\ [\ref{eq_tw}] and [\ref{eq_ra}];
discussed in \S \ref{sec_results}).  Overall, the magnetic cases
presented in section \ref{sec_results} resulted in stellar spin
periods in the range of approximately 1--10 days, in the age range of
1--3 Myr.  This range covers the range of the most populated regions
of the observed spin period distributions of clusters with similar
ages \citep[e.g., ][]{stassunea99, herbstea02, rebull3ea04, lammea05,
  herbstea07, irwinbouvier09}.  In general, accretion-powered stellar
winds may explain the slowest rotators as those stars with the
strongest magnetic fields, lowest accretion rates, and/or highest
stellar wind mass outflow rates (relative to the accretion rate).

We found that some cases that had torques strong enough to drive the
system close to an equilibrium spin rate (e.g., as in Fig.\
\ref{fig_10b2000}).  Spin equilibrium states are interesting because
the initial conditions are ``erased,'' and the spin rate depends only
on present conditions (e.g., accretion rate, magnetic field strength,
etc.).  However, for the conditions considered here (in particular,
for strong magnetic coupling), the magnetic spin-down torques arising
in the star-disk interaction were completely negligible.  Thus, these
stars' equilibrium rotation rates are not physically linked to the
rotation rate of the disk inner edge (as in disk locking) but rather
are described by a simple balance between a spin-up torque from
accretion and a spin-down torque from a stellar wind.  This simple
balance allows for an analytic formula for predicting the equilibrium
(net-zero torque) spin rate (e.g., eq.\ [\ref{eq_feq}]).

While we have necessarily made a number of simplifying assumptions, we
have whenever possible adopted the same assumptions that are typically
used in previous models in the literature (in particular, the
disk-locking models).  Our approach in Paper~I and here has been
deliberately systematic, so that each additional component or
assumption in the model can be understood in turn.  In this way, our
approach best serves to highlight the influence of the opening of the
mangetic field due to star-disk differential rotation (Paper~I) and
the influence of an acctetion-powered stellar wind (present paper),
{\it relative} to the previous studies that do not include these
effects.  We have explored a range of conditions that are consistent
with observations of T Tauri stars, but we have not explored all of
parameter space, and there remain uncertainties inherant to various
adopted assumptions and approximations.  Thus, while our quantitative
results should be viewed as approximate, the results relative to other
studies/models in the literature are robust.

With this in mind, we compared the predicted equilibrium spin rate of
the APSW scenario to the predictions of two types of disk-locking
models, the Ghosh \& Lamb type and X-wind models (\S
\ref{sec_comparison}).  Overall, this comparison and our results in
general demonstrate that APSWs can explain the observed distribution
of young star spins in a similar way as the classical disk locking
picture, while at the same time avoiding the problem of magnetic field
line opening (for the GL-type models), as well as the assumption of
spin equilibrium and requirement of significant flux trapping (for the
X-wind model).

The APSW scenario has one additional parameter, the mass loss rate in
the wind.  For the assumptions adopted here, we found that in order
for the APSWs to have a significant influence, the mass loss rates
should be at least of the order of a percent of the accretion rate.
It still remains to be shown whether and/or how the energy derived
from the accretion process may drive a wind that is magnetically
connected to the star, and with sufficiently high mass outflow rate.
Such winds would not likely be driven significantly by thermal
pressure \citep[due to a rapid cooling time][]{mattpudritz07iau}, nor
by magneto-centrifugal effects (from slowly-rotating stars), but
Alfv\'en waves may be important \citep[e.g.,][]{Decampli:1981p2972,
  Hartmann:1982p2885, suzuki07}.  \citet{Cranmer:2008p1657,
  Cranmer:2009p1647} demonstrated that Alfv\'en waves generated by the
accretion process in T Tauri stars are capable of driving enhanced
stellar winds.  The mass loss rates derived by those models were
typically $\chi \la 0.01$.  There is a hard energetic upper limit of
$\chi \la 0.6$ for APSWs \citep[][]{mattpudritz08III}, and values
close to this will have observational consequences that may already be
ruled out in some systems \citep{zanniferreira11}.  Thus, values of
$\chi$ much greater than 10--20\% seem unlikely, in general.  Clearly,
more work is needed to determine what is the mass loss rate along the
stellar magnetic field and how this should depend upon system
parameters.

Measurements of the magnetic field strengths of young stars suggest
that the stellar surface is blanketed with complex magnetic fields
with a strengths of $\sim 2$~kG \citep[e.g,.][]{safier98,
  Bouvier:2007p3031, johnskrull07, yangjohnskrull11}.  However, even
when complex magnetic fields are present, it is the large-scale
(dipole) component that is the most important for the star-disk
interaction and torques on the star \citep[e.g.,][]{gregoryea08}.  The
dipole components have been measured via spectropolarimetry for at
least 10 accreting T Tauri stars to date \citep{donatiea07v2129oph,
  donatiea08bptau, hussainea09, donatiea10aatau, donatiea10v2247oph,
  donatiea11twhya, donatiea11v2129oph, donatiea11v4046sgr,
  skellyea11mtori}, and the equatorial field strengths of the dipole
components range from $\sim 100-1000$ G, with the top of this range
determined by just two of the stars ($B_*\approx 600$~G for BP Tau,
and $B_*\approx 1000-1500$~G for AA Tau ; \citealp{donatiea08bptau,
  donatiea10aatau}).  The disk-locking models usually require field
strengths of $\sim 1000$G to explain the slowest rotators, and the
weakest equatorial field strength we have considered in the present
work is $B_*=500$G.  The APSW scenario can, in principle, make up for
a weaker magnetic field by having a larger mass outflow rate, although
the largest values of $\chi$ considered in the present paper are
already approaching the upper limits \citep[see,
e.g.,][]{mattpudritz08III, zanniferreira11}.  To better constrain the
models, it will be important to have magnetic field measurements,
particularly of the large-scale components, for a larger sample of
pre-main-sequence stars.

In addition to those discussed above, there remain a number of caveats
to the present work.  While we have adopted a relatively sophisticated
model for the torques on the star, we use a simplified treatment for
the accretion history, the evolution of the magnetic field, and the
structure and evolution of the star itself.  In order to be able to
make a more meaningful comparison with observations, and especially to
be able to evolve the system to much later times, it will be necessary
to improve the theoretical treatment of these components.

For example, the present treatment of the accretion does not allow for
a possible state of the system in which the disk is truncated outside
of the corotation radius, the so called ``propeller'' regime.  In this
regime, there is generally no accretion onto the star
\citep[e.g.,][]{illarionovsunyaev75, Sunyaev:1977p3829}, or the
accretion is intermittent \citep[e.g.,][]{romanovaea05,
  DAngelo:2010p3127}.  In the present work, we are considering systems
with strong magnetic coupling (large magnetic Prandtl number in the
disk) and relatively slow rotation.  Under these conditions, the
magnetic spin-down torque on the star---which acts to spin up the disk
and can potentially truncate the disk outside of $R_{\rm co}$---is
relatively weak compared to other torques in the system.  Thus, we
don't expect the propeller regime to be important for most of the
evolution these systems.  However, there may be times in the history
of such stars (e.g., toward the end of the accretion phase) in which
the propeller regime is important for the angular momentum loss.
Properly capturing the transition between accreting and propeller
states requires a self-consistent treatment of the evolution of the
accretion disk \citep[as in, e.g.,][]{armitageclarke96,
  DAngelo:2010p3127}.  In the present model, we imposed the accretion
rate onto the star, which implicitly assumes that the disk is always
able to transport excess angular momentum given to it by the star
\citep[see, e.g.,][]{mattpudritz05}.  To explore the influence of the
propeller regime on the long-term spin evolution, future models should
include both self-consistent disk evolution and sophisticated
treatment of the torques on the star.

Finally, all of our conclusions regarding the effectiveness of stellar
wind torques relies on the particular formulation for the torque from
\citet{mattpudritz08II}.  While this is the most appropriate
formulation for low-mass-star winds that exists in the literature,
there remain a few open questions about its use in the present work.
First of all, that formulation was derived for an isolated star, and
we have simply added this torque to our model, which also assumes the
presence of an accretion disk and associated interaction torques in a
non-self-consistent way.  It is not yet clear how the presence of a
disk will influence the stellar wind, and in particular how it may
influence the dependence of the stellar wind torque on parameters
(e.g., the value of $m$).  Also, the formulation of
\citet{mattpudritz08II} was based upon simulations of a star with a
singular spin rate ($f=0.1$) and a solar-like wind acceleration
mechanism.  It is encouraging that a study of angular momentum flow
from massive stars \citep{uddoula3ea08, uddoula3ea09}, which includes
a very different wind acceleration mechanism and a range of spin
rates, found a similar power law relationship for the torque, but more
studies are warranted.  Lastly, all stellar wind torque formulations
in the literature are based upon simple (e.g., dipolar) magnetic
geometries, while we know that T Tauri stars possess complex magnetic
field structures.  In future work, it will be important to determine
whether the presence of an accretion disk, more complex magnetic
geometries, or different wind acceleration mechanisms will
significanlty enhance or suppress the stellar wind torque relative to
the formulation of \citet{mattpudritz08II} adopted here.

Much work remains to develop the necessary theory, to improve the
precision and number of observational measurements of relevant system
parameters, and ultimately to understand the observed distributions
and evolution of stellar spin rates, but the idea that powerful
stellar winds may extract significant angular momentum from accreting
stars remains a promising scenario.

\acknowledgments

We thank the anonymous referee for useful remarks on the manuscript.
SPM was supported by an appointment to the NASA Postdoctoral Program
at Ames Research Center, administered by Oak Ridge Associated
Universities through a contract with NASA, and by the ERC through
grant 207430 STARS2 (http://www.stars2.eu).  TPG acknowledges support
from NASA's Origins of Solar Systems program via WBS
811073.02.07.01.89.



\end{document}